\documentclass[fleqn,usenatbib]{mnras}
\usepackage{newtxtext,newtxmath}
\usepackage[T1]{fontenc}
\DeclareRobustCommand{\VAN}[3]{#2}
\let\VANthebibliography\thebibliography
\def\thebibliography{\DeclareRobustCommand{\VAN}[3]{##3}\VANthebibliography}
\usepackage{graphicx}	
\newcommand{\kepler}{{\sl Kepler}\ }
\title[Kepler-1660 Circumbinary Planet]{A $5M_{\rm Jup}$ Non-Transiting Coplanar Circumbinary Planet Around Kepler-1660AB}

\author[M. Goldberg et al.]{
Max Goldberg,$^{1,2}$\thanks{Email: mg@astro.caltech.edu}, 
Daniel Fabrycky,$^{2}$ 
David V. Martin,$^{3,4}$
Simon Albrecht,$^{5}$
Hans J. Deeg,$^{6,7}$
\newauthor
and Grzegorz Nowak$^{6,7}$
\\
$^{1}$Department of Astronomy, California Institute of Technology, 1200 E. California Blvd, Pasadena, CA 91125, USA\\
$^{2}$Department of Astronomy \& Astrophysics, University of Chicago, Chicago, IL 60637\\
$^{3}$Department of Astronomy, The Ohio State University, 4055 McPherson Laboratory, Columbus, OH 43210, USA\\
$^{4}$NASA Sagan Fellow\\
$^{5}$Stellar Astrophysics Centre, Department of Physics and Astronomy,
Aarhus University, Ny Munkegade 120, DK-8000 Aarhus C, Denmark\\
$^{6}$Instituto de Astrof\'\i sica de Canarias, C. V\'\i a L\'actea S/N, E-38205 La Laguna, Tenerife, Spain\\
$^{7}$Universidad de La Laguna, Dept. de Astrof\'\i sica, E-38206 La Laguna, Tenerife, Spain
}

\pubyear{2023}

\begin{document}
\label{firstpage}
\pagerange{\pageref{firstpage}--\pageref{lastpage}}
\maketitle

\begin{abstract}

Over a dozen transiting circumbinary planets have been discovered around eclipsing binaries. Transit detections are biased towards aligned planet and binary orbits, and indeed all of the known planets have mutual inclinations less than $4.5^{\circ}$. One path to discovering circumbinary planets with misaligned orbits is through eclipse timing variations (ETVs) of non-transiting planets. \citet{Borkovits2016} discovered ETVs on the 18.6 d binary Kepler-1660AB, indicative of a third body on a $\approx 236$ d period, with a misaligned orbit and a potentially planetary mass. \citet{Getley2017} agreed with the planetary hypothesis, arguing for a $7.7M_{\rm Jup}$ circumbinary planet on an orbit that is highly misaligned by $120^{\circ}$ with respect to the binary. In this paper, we obtain the first radial velocities of the binary. We combine these with an analysis of not only the ETVs but also the eclipse depth variations. We confirm the existence of a $239.5$ d circumbinary planet, but with a lower mass of $4.87M_{\rm Jup}$ and a coplanar orbit. The misaligned orbits proposed by previous authors are definitively ruled out by a lack of eclipse depth variations. Kepler-1660ABb is the first confirmed circumbinary planet found using ETVs around a main sequence binary.

\end{abstract}


\section{Introduction} \label{sec:intro}

The discovery of planets orbiting around two stars -- circumbinary planets -- has provided a new perspective on planet formation and evolution. The \kepler mission led to the discovery of 14 transiting circumbinary planets (reviews in \citealt{Welsh2018,Martin2018a}). These circumbinary systems are highly reliable, owing to a transit signature with variable depth, duration and timing that has no known false-positive mimickers \citep{Armstrong2013,Kostov2014,Windemuth2019,Martin2021}. Despite a small sample, we are beginning to uncover trends in this population. Of particular interest is the alignment between the binary and planet orbital planes. The distribution of mutual inclinations has implications for disc alignment and processes such as scattering and Kozai-Lidov which may misalign planetary orbits \citep{Martin2015a}.

A challenge with transits is that they are biased to systems with aligned planet and binary orbital planes \citep{Martin2014,Chen2022}. All of the known transiting circumbinary planets reside on orbits within $4.5^{\circ}$ of coplanarity with the binary orbit. \citet{Martin2014,Armstrong2014} showed that a coplanar distribution equates to a similar abundance of gas giants around one and two stars. This implies that a hidden population of circumbinary planets on misaligned orbits would be indicative of a greater abundance of giants around binaries. The likelihood of such a surprising scenario has been reduced by the transit studies of \citet{Li2016,Chen2022} and the radial velocity work of \citet{Martin2019a}, all of which constrain an inclination distribution that is largely flat. However, this does not rule out that some outlier misaligned circumbinary planet orbits may exist. 

One method to detect non-transiting circumbinary planets regardless of inclination is to use Eclipse Timing Variations (ETVs). For ETVs there are two classes of discovery, corresponding to types of systems. The first is planets around post-common envelope binaries, for which NN Ser \citep{Qian2009} was the first of over a dozen proposed planets \citep{Pulley2022}. These very tight ($P_\text{bin}<1$ d) binaries contain at least one evolved star. The ETVs are caused by the light travel time effect, which grows with the planet period. Consequently these are typically long-period ($\sim$ several au) planets. However, most of these discoveries have been disputed for a variety of reasons, including  orbital stability \citep{Wittenmyer2013,Horner2013}, confusion with magnetic activity cycles (\citealt{Bours2016}, including the Applegate mechanism \citealt{Applegate1992}), low statistical significance \citep{Hinse2014} and direct imaging disproval \citep{Hardy2015}. An argument from \citet{Zorotovic2013} is that $\sim90$ per cent of post common envelope binaries have ETVs, which would imply an unrealistically large planet abundance.

The second discovery class is dynamical ETVs, where the third body directly perturbs the binary's Keplerian orbit. For the transiting circumbinary planets discovered by \kepler, these ETVs were used in about half of the cases to confirm the planet and measure its mass. In the remaining sample, the lack of ETVs constrains the mass of the third body to be planetary. In terms of discovering new, non-transiting planets using ETVs, the most systematic search has been \citet{Borkovits2016} using \kepler eclipsing binaries. One particularly interesting candidate they identified was KIC 5095269, for which the \citet{Borkovits2016} fit suggested a tertiary body with $P_\text{P}=236.26\pm0.8 $ d and $m_\text{P}=19\pm8M_\text{Jup}$ and a misaligned orbit with $i_m=40^{\circ}\pm1^{\circ}$. This would suggest a circumbinary brown dwarf or high mass planet on a misaligned orbit.

\citet{Getley2017} followed-up with a dedicated study of KIC 5095269. Using the same \kepler photometry and an \textit{N}-body model, they recovered a planet with similar period but lower mass $M_\text{P}=7.70\pm0.08M_\text{Jup}$ and a higher mutual inclination $i_m = 120^{\circ}$. The binary was found to have period $P_\text{bin}=18.61196$ days and stellar masses $M_\text{A}=1.21M_{\odot}$ and $M_\text{B}=0.51M_{\odot}$, as determined by eclipse depths and a $V-K$ color.

If confirmed, KIC 5095269 would represent several firsts: the first ETV discovered planet around a main sequence binary, the most massive circumbinary planet and the first on a highly misaligned orbit. \cite{Getley2017} demonstrate that the misaligned orbit is stable, using \textit{N}-body integrations over $10^7$ yr. This is expected since the period ratio $P_{\rm P}/P_{\rm bin}=12.7$ is greater than almost all of the \kepler systems and exceeds the numerical stability limits of \cite{Holman1999}. It has also been shown by multiple studies that circumbinary planets  may be stable at any mutual inclination \citep{Farago2010,Doolin2011,Martin2016}. There are also plausible mechanisms for producing misaligned planetary orbits, based on observations  \citep{Kennedy2012,Kennedy2019} and theory \citep{Martin2019,Childs2021,Rabago2023} of polar circumbinary discs, planet-planet scattering \citep{Smullen2016} or interactions with a third star \citep{Munoz2015,Martin2015a,Hamers2016}.

Therefore, while the \citet{Getley2017} model is theoretically plausible, more observations and analysis are required to confirm the existence of a planet. Since ETVs yield an indirect detection of the planet, our knowledge of the planet is dependent on how well the binary is characterized. Unusually, Kepler-1660AB only has primary eclipses, and the \citet{Borkovits2016,Getley2017} studies did not have the benefit of radial velocities.\footnote{Indeed, \citet{Getley2017} encouraged radial velocity measurements to be taken.} The binary eccentricity, masses and mass ratio were therefore poorly constrained. Furthermore, the analysis of \citet{Getley2017} also only accounted for variable eclipse \textit{timing} and not variable eclipse \textit{depths}, the latter of which could be indicative of a changing binary inclination due to a massive, highly misaligned tertiary body.
 
In this paper we conduct a new analysis of KIC 5095269, including 12 radial velocity measurements taken with the CARMENES and FIES spectrographs. We simultaneously fit the eclipse times, eclipse depths, and radial velocities. Our fits allow for all possible orbital configurations, including special tests for the misaligned solutions from \citet{Borkovits2016} and \citet{Getley2017} solution, and a polar solution predicted theoretically by \citet{Martin2017} and \citet{Farago2010}. Overall we confirm that the there is indeed a circumbinary planet in KIC 5095269, which has subsequently been named Kepler-1660ABb. The planet has a period of 239 days and mass of $4.87M_{\rm Jup}$, which is less massive but qualitatively similar to the \citet{Getley2017} model. However, we robustly rule out a misaligned solution, largely based on constant eclipse depths over the four year \kepler baseline. We constrain Kepler-1660ABb to be coplanar within $6.4^\circ$, like the known transiting planets.

In Section \ref{sec:data}, we describe the data and methods.  In Section \ref{sec:results}, we present the results. We discuss our findings in Section \ref{sec:discussion} in the context of previous work, the formation and stability of the planet and future observations. We ultimately conclude in Section \ref{sec:conclusion}.

\section{Data and Methods}
\label{sec:data}

\subsection{Kepler data}

Kepler-1660AB was determined to be a detached eclipsing binary with just the first 44 days of \kepler data \citep{Prsa2011}. Only one set of eclipses is seen; we will call the star that is covered during the eclipses the `primary,' regardless of whether it is the more massive star. 

The photometric data are the long-cadence SAP fluxes recorded by the \kepler team at the MAST data archive\footnote{\url{https://archive.stsci.edu/}}. After masking eclipses, each quarter of data was detrended using the 5 most significant cotrending basis vectors. Data with quality flags (SAP\_Quality$\geq 16$) indicating problems were excluded from the analysis. 

Our analysis of the photometry starts with computing the shape of the eclipse. We model the relative position of the stars near eclipse as rectilinear motion in time. This produces a mid-time in which the stars are closest to one another, and a transit duration from first to fourth contact. The flux is modeled with the \cite{Mandel2002} \textsc{occultnl} code in \textsc{IDL}, diluted by the (assumed constant) fractional flux of the secondary, $F_s$. The ratio of radii $R_2/R_1$ and the limb-darkening coefficient $u_1$ (linear limb-darkening was used) are also free parameters. Each transit mid-time is a free parameter, so that eclipse timing variations are also a result of this fit. The SAP flux data are divided by the model, then a third-order polynomial is fit to the residuals, to take into account instrumental and astrophysical variability. Our measured eclipse times are given in Table~\ref{tab:ecl}.

We also produced a clean version of the SAP lightcurve by masking the eclipses, fitting the first 5 cotrending basis vectors and subtracting them as an instrumental model. Then, we further detrend the astrophysical variability using a 3rd order polynomial fit to 500 minutes before and after each datapoint as a model for what it should be. Those polynomials interpolate the eclipse fluxes. Hence the eclipses are unmasked on a detrended background. 

\begin{figure*}
    \centering
    \includegraphics[width=\textwidth]{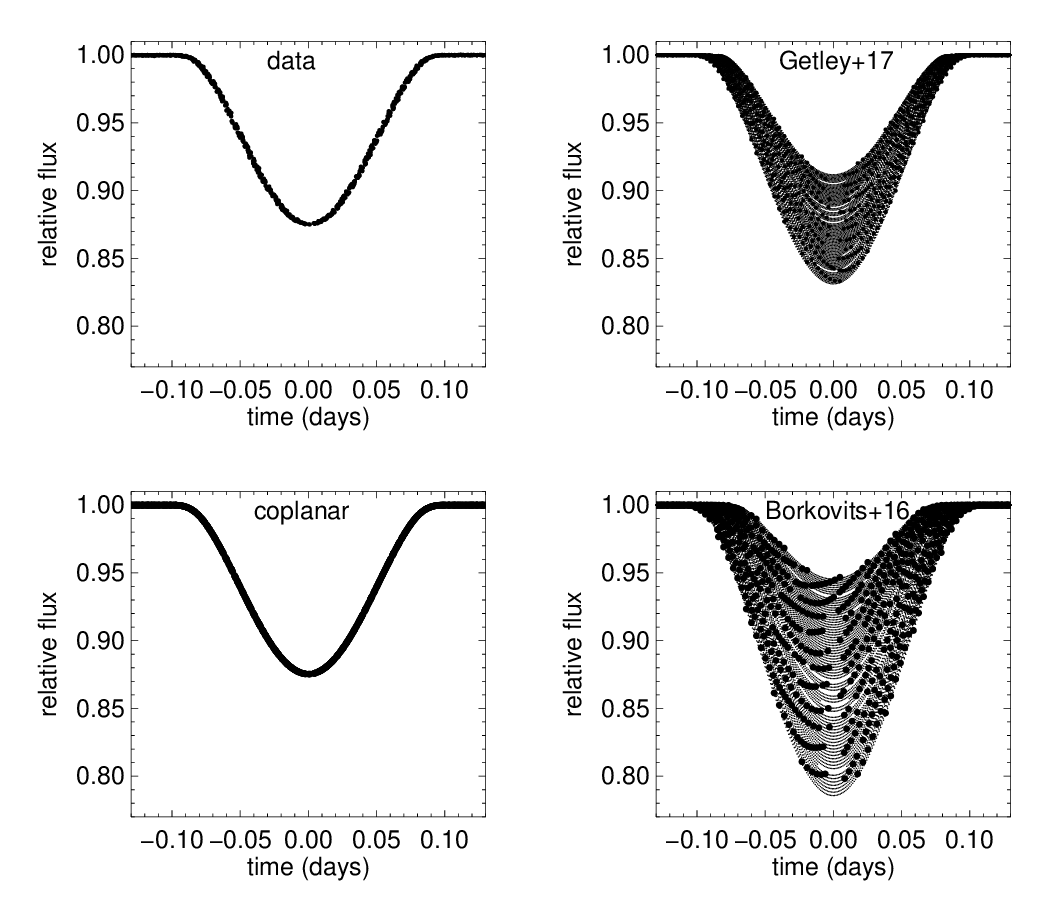}
    \caption{Changes to the lightcurves induced by a massive and inclined planet. The top-left panel shows the detrended \kepler data near and in primary eclipse; the small horizontal scatter in the ingress and egress is due to the ETVs induced by the planet. The bottom-left panel is a fit using \textsc{occultnl} and a circular orbital model of a dark secondary with an imposed impact parameter of 1.2. A linear-in-time change to the inclination is allowed in the fit, and is found to be $9.9\times10^{-5} {\rm deg yr}^{-1}$. (Later we shall see that a dark secondary on a circular orbit is not realistic, and eccentricity explains the lack of secondary eclipses. This plot merely quantifies the inclination change using a model agreeing with the lightcurve shape.) An \textit{N}-body model of the orbital parameters of \citet{Getley2017} and \citet{Borkovits2016} has inclination changes of $0.0720 {\rm deg yr}^{-1}$ and $-0.144 {\rm deg yr}^{-1}$, which produces a very large depth increase and decrease, respectively (top-right and bottom-right) when other model parameters are held fixed from our previous fit ($b=1.2$, $R_B/R_A=0.723$; $\rho_{\rm {A,circ}} = 3.76514 {\rm g cm}^{-3}$; $u_1=0.678$). In these two panels, instantaneous lightcurve values are shown as small dots, and those are binned on the long-cadence timescales to the observable larger points.} 
    \label{fig:compare}
\end{figure*}

The planetary orbits suggested by \cite{Borkovits2016} and \cite{Getley2017} are unusual in that the mutual inclination $i_m$ is far from a multiple of $90^\circ$. This misalignment induces a precession on the inner binary. While only a fraction of a degree, any change in the binary inclination should be readily apparent in the eclipse depths derived from \kepler data. Over the $\sim1500$ d \kepler baseline, a third body in the orbit given by \cite{Borkovits2016} with $i_m = 40^\circ$ would cause the binary inclination to change by $-0.57^\circ$, and a planet in the orbit given by \cite{Getley2017} with $i_m = 120^\circ$ (indicating a retrograde orbit) would cause the binary inclination to change by $+0.28^\circ$. In Fig.~\ref{fig:compare}, we compare the phase-folded \kepler data (top left) with a lightcurve model of three orbital configurations. The highly misaligned solutions from \citet{Getley2017} (top right) and \citet{Borkovits2016} (bottom right) lead to strong changes in the eclipse depth not visible in the observations.

To extend this argument quantitatively, we include a linear drift in the impact parameter in the photometric fit. Specifically, we define the instantaneous impact parameter to be
\begin{equation}
    b(t) = b_{0,\rm{obs}} + db/dt_{\rm obs} \cdot t
\end{equation}
where $t$ is the time since BJD-2545900. We find $b_{0,\text{obs}} = 1.30 \pm 0.20$ and $db/dt_\text{obs} = (3.99 \pm 1.69)\times 10^{-7}$ d$^{-1}$, corresponding to a precession rate three orders of magnitude slower than predicted by \cite{Getley2017} and \cite{Borkovits2016}. This strict limit on the precession of the binary is incorporated in the dynamical model as detailed below.

\cite{Getley2017} argued that the lack of secondary eclipses in the \kepler data suggests the secondary star is much less massive and luminous than the primary. We argue that since the \kepler photometry is so precise, secondaries of almost any depth would have been noticed. In fact, with our radial velocities we will eventually conclude that the stars are very similar in mass and radius. Instead, the lack of secondary eclipses is due to a particular geometry of the orbits that leads to overlap of the stellar discs at one conjunction and not the other.

For this to happen, the orbital eccentricity must be non-zero and the orbit must be inclined to the line of sight.  We define the argument of periapse $\omega$ to be the angle between the sky plane and the periapse of the secondary's orbit about the primary; then primary eclipse occurs at a true anomaly $f$ of $90^\circ-\omega$. The distance between the stars at that moment is then $r_1 = a (1-e^2) / (1+e \sin \omega)$, where $a$ and $e$ are the semi-major axis and eccentricity of the eclipse traced by the secondary's position with respect to the primary. The inclination (defined with $0^\circ < i <90^\circ$) must be close enough to $90^\circ$ for eclipses to occur.  Yet it must not be too close, because when the primary is closer to the observer, the distance between the two stars is $r_2 = a (1-e^2) / (1-e \sin \omega)$, and then the stars must not cross.  Together, we have the constraint on $i$: 
\begin{equation}
\frac{R_1+R_2}{a} \frac{1-e \sin \omega}{1-e^2} < \cos i < \frac{R_1+R_2}{a} \frac{1+e \sin \omega}{1-e^2},
\label{eq:bin_const}
\end{equation}
and for this inequality to be possible, $0^\circ < \omega < 180^\circ$. 

\subsection{Spectroscopic Data}

We obtained high-resolution echelle spectra of Kepler-1660 with two different instrument/telescope combinations. These spectra allow us to unambiguously show that the two stars in Kepler-1660 have very similar masses and luminosities, in stark contrast to the photometry-only results of \cite{Getley2017}.
We processed each spectrum to obtain the radial velocity (RV) of each star at every epoch. The recovered RVs are given in in Table~\ref{tab:RVs} and included in the dynamical model in Section~\ref{sec:dyn}.

\subsubsection{FIES}\label{subsubsec:fies}

We observed Kepler-1660 with the Fiber-Fed Echelle Spectrograph \citep[FIES;][]{Telting2014} mounted to the Nordic Optical Telescope with its $2.56$ m  primary mirror during the summer and fall of 2017. FIES covers the wavelength range from $370$ to $830$\,nm and has in the high resolution mode, employed by us, a spectral resolution of $R\approx 67\,000$. All science exposures were 3000\,s and accompanied by a 600\,s ThAr calibration exposure to determine the wavelength scale for that particular observation. We used FIEStool \citep{Telting2014} for the data reduction to obtain a final spectrum.
The typical signal-to-noise ratio (SNR) per observation was 2 -- 10.

\subsubsection{CARMENES}\label{subsubsec:carmenes}

We later collected data on the CARMENES spectrograph \citep{Quirrenbach2014} on the Calar Alto 3.5 m telescope in the spring of 2018. The spectrograph covers the wavelength range 520 -- $960$\,nm at resolution $R\approx 82\,000$. Exposures were 900 seconds each. We cross-correlated the spectrum (see below) in the wavelength range 550 -- 680\,nm.

\begin{figure}
    \centering
    \includegraphics[width=0.5\textwidth]{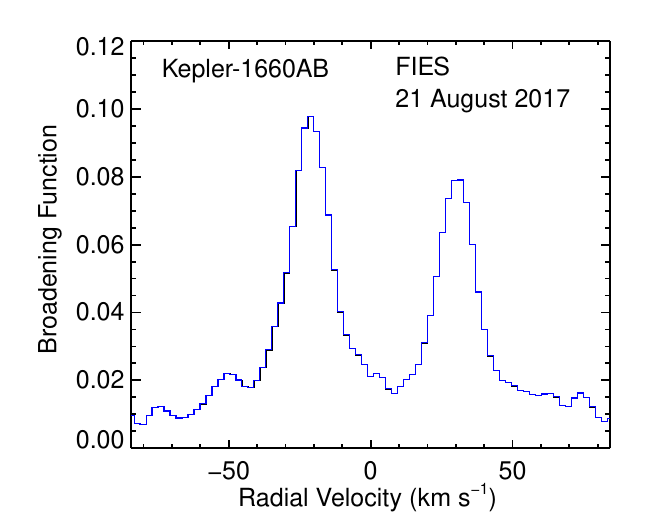}
    \caption{The broadening function \citep{Rucinski1999} calculated for one of our spectroscopic observations. Two clear peaks are visible, indicating a double-lined spectroscopic binary (SB2).}
    \label{fig:spec}
\end{figure}

\subsubsection{Radial Velocity Measurements}
Many of the spectra showed two sets of lines, indicating a double lined spectroscopic binary (SB2) with nearly equal mass components.

To obtain RVs from the FIES data, we used the Broadening Function technique \citep[BF;][]{Rucinski1999} on each spectral order of each of the reduced spectra. We used a Phoenix spectrum \citep{Husser2013} with $T_{\rm eff}=6500$~K, $\log g=4.5$ and [Fe/H] $=0.0$. In the BFs two peaks are visible with similar heights and areas. This indicates that the two stars are similar in mass, in contrast to \citet{Getley2017}, who derived unequal stellar masses without the benefit of high-resolution spectroscopy. One example BF is plotted in Fig.~\ref{fig:spec}. 
For each spectral order, we fitted a double-Gaussian model and the position of the centroids of the Gaussians were then taken as RVs of the two stellar components. For each epoch, the final RVs measurements and their uncertainties were obtained from the SNR-weighted RV average and scatter obtained for the different orders.

For the CARMENES data, a TODCOR \citep{Zucker1994} implementation\footnote{IDL program by James Davenport: \url{https://github.com/jradavenport/jradavenport_idl/blob/master/pro/todcor.pro} } was used to determine the RVs. 
We cross-correlated the data against the HARPS.GBOG\_Procyon.txt and HARPS.GBOG\_HD84937-1.fits templates from \cite{Blanco-Cuaresma2014}. While the template stars have different temperatures, the fit was similar if their velocities were swapped, so we assign the output velocities to each star to result in the most coherent Keplerian motion. We set the $1\sigma$ uncertainty to 4 km/s for each CARMENES RV so that the reduced $\chi^2$ of the orbital fit is approximately unity.

\subsection{Dynamical Model}
\label{sec:dyn}
We modeled the eclipse times and radial velocities of Kepler-1660 with a three-body model using \textsc{rebound}, an \textit{N}-body integrator in Python \citep{Rein2012}. The code repository is publicly available.\footnote{\url{https://github.com/goldbergmax/etv-fit}} We used the IAS15 algorithm with adaptive step size, which is generally accurate to machine precision \citep{Rein2015a}. Our algorithm assumes that mid-transit times occur when the projected separation is perpendicular to the projected velocity. Near expected eclipse times, our algorithm solves the equation $\textbf{x} \cdot \textbf{v} = 0$ using the Newton-Raphson method until the step size becomes less than $10^{-10}$ d. The eclipse time is corrected for the Light Travel Time Effect (LTTE) \citep{Borkovits2016} by adding the light travel time from the radial center-of-mass of the binary stars to the radial center-of-mass of the entire system. In practice, the LTTE ETVs have an amplitude $\sim 30$ times smaller than the dynamical ETVs.

The model has 14 free parameters. All parameters are osculating values at the epoch BJD-2454900. Two parameters, $P_1$ and $P_2$, encode the periods of the secondary star and planet, respectively. Three parameters, $M_A$, $M_B$, and $M_p$, give the masses of the primary, secondary, and planet. The eccentricity and argument of periapse are parametrized by $e_1$ and $\omega_1$ for the binary and $e_2\cos\omega_2$ and $e_2\sin\omega_2$ for the planet. The sky-plane inclinations are $i_1$ and $i_2$. The longitudes of the ascending node are $\Omega_1$ and $\Omega_2$, but because the data only constrain $\Delta \Omega = \Omega_2 - \Omega_1$, we set $\Omega_1=0$ during the fitting.\footnote{The \citet{Getley2017} fit contains non-zero values for both $\Omega_1$ and $\Omega_2$, but like ours, their fit is only sensitive to $\Delta \Omega$.} The phase of the binary is represented by $T_{01}$, the time of inferior conjunction assuming no planetary perturbation of the binary orbit. This time would correspond to the time of the first eclipse for a sufficiently edge on orbit, and is well-constrained by the data. The orbital phase of the planet is represented by $T_{\rm p}$, the first time after BJD-2454900 that the planet passes through periapse. The orbital elements of the secondary star are defined relative to the primary star, and those of the planet are defined relative to the center-of-mass of the primary and secondary stars. Finally, to account for the radial velocity of the system's barycenter and because our radial velocity data were reduced separately, we include two radial velocity zero point offsets, $\gamma_1$ and $\gamma_2$ for the NOT and CARMENES data, respectively.

The model outputs eclipse times, individual stellar radial velocities at predetermined times, and the impact parameter at each eclipse. To match the simulated eclipse impact parameters to the observed $b_{0,\text{obs}}$ and $db/dt_\text{obs}$, we fit the simulated impact parameters to the linear model $b(t) = b_{0,\text{sim}} + db/dt_\text{sim}\cdot t$, where $t$ is given in BJD-2454900.

For the eclipse times, we define the goodness-of-fit statistic
\begin{equation}
    \chi^2_\text{ETV} = \sum_i \left(\frac{E_{\text{mod},i} - E_i}{\sigma_i}\right)^2
\end{equation}
where $E_i$, $E_{\text{mod},i}$, and $\sigma_i$ are the observed eclipse midpoint, modeled eclipse midpoint, and timing uncertainty of the $i$-th eclipse, respectively.

For the RV measurements of stars A and B, we define the goodness-of-fit statistic 
\begin{equation}
    \chi^2_\text{RV} = \sum_j \left(\frac{RV_{\text{A,mod},j} - RV_{\text{A},j}}{\sigma_{\text{A},j}}\right)^2 + \left(\frac{RV_{\text{B,mod},j} - RV_{\text{B},j}}{\sigma_{\text{B},j}}\right)^2
\end{equation}
where  $RV_j$, $RV_{\text{mod},j}$, and $\sigma_j$ are the observed radial velocity, modeled radial velocity, and radial velocity uncertainty of the $j$-th measurement, respectively.

Finally, for the eclipse depth measurements, we define the goodness-of-fit statistic
\begin{equation}
    \chi^2_{b(t)} = \left(\frac{db/dt_\text{sim} - db/dt_\text{obs}}{\sigma_{db/dt}}\right)^2 + \left(\frac{b_{0,\text{sim}} - b_{0,\text{obs}}}{\sigma_{b_0}}\right)^2
    \label{eq:chisq_b}
\end{equation}
where $db/dt_\text{obs}$ and $b_{0,\text{obs}}$ are the observed eclipse impact parameter change slope and constant with their respective uncertainties $\sigma_{db/dt}$ and $\sigma_{b_0}$, given in Section 2.1, and $db/dt_\text{sim}$ and $b_{0,\text{sim}}$ are the simulated eclipse impact parameter change slope and intercept.

From these statistics, we adopt a total log-likelihood of
\begin{equation}
    \log \mathcal{L} = -\frac{1}{2} \chi^2_\text{ETV} - \frac{1}{2} \chi^2_\text{RV} - \frac{1}{2} \chi^2_{b(t)}.
\end{equation}
We also set $\log \mathcal{L} = -\infty$ for samples that did not satisfy Eq.~\ref{eq:bin_const}, i.e. there must be a primary eclipse but not a secondary one.

The Affine-Invariant MCMC Ensemble sampler from \texttt{emcee} was used to sample the parameter space and determine parameter uncertainties. We evolved 100 chains for 40\,000 generations and removed the first 5000 as burn-in.

\subsection{Isochrones Model}
While eclipsing binaries usually provide an opportunity to measure absolute stellar radii, in this case the lack of secondary eclipses constrains only the ratio of radii. However, the absolute stellar masses derived from radial velocities can be used along with photometric fluxes and parallax measurements to estimate stellar radii. We modeled the pair of stars using the \texttt{isochrones} package and its included MIST stellar models \citep{Morton2015}. Because the stellar masses derived from the radial velocity fit are correlated, we parametrized the stellar mass prior as a bivariate normal distribution with means $M_A,M_B$ and correlation $\rho_M$ given in Table \ref{tab:phot}.
We also include a variety of photometric and astrometric parameters, detailed in Table \ref{tab:phot}. The model parameters were masses for each star, and a joint metallicity, age, distance, and \textit{V} -- band extinction. We find a primary star radius of $R_A = 1.49 \pm 0.07 R_\odot$ and a secondary star radius of $R_B = 1.30 \pm 0.06 R_\odot$. We use these values, without uncertainties, in the dynamical model to compute the impact parameter of each eclipse. The fit also recovers effective temperatures of the primary and secondary stars of $6050 \pm 40$\,K and $6040 \pm 30$\, K as well as a primary-to-secondary flux ratio of $1.32\pm 0.22$.

\begin{table}
\centering
\caption{The photometric and astrometric measurements used in the \texttt{isochrones} fit.
\label{tab:phot}}
\begin{tabular}{cccc}
Parameter & Value & Uncertainty & Source \\
\hline
Parallax (milliarcsec) & 0.8126 & 0.0148 & Gaia DR2\\
G magnitude & 13.4394 & 0.0002 & Gaia DR2\\
BP magnitude & 13.7217 & 0.0017 & Gaia DR2\\
RP magnitude & 12.9942 & 0.0009 & Gaia DR2\\
J magnitude & 12.499 & 0.023 & 2MASS \\
H magnitude & 12.217 & 0.018 & 2MASS \\
K magnitude & 12.215 & 0.024 & 2MASS \\
$M_A$ ($M_\odot$) & 1.1366 & 0.0728 & This work \\
$M_B$ ($M_\odot$) & 1.0820 & 0.0644 & This work \\
$\rho_M$ & 0.94504 & -- & This work \\
$R_A$ ($R_\odot$) & 1.491 & 0.072 & This work \\
$R_B$ ($R_\odot$) & 1.299 & 0.061 & This work
\end{tabular}
\end{table}

\section{Results}
\label{sec:results}
The best-fitting solution and uncertainties derived from our MCMC chains are shown in Table \ref{tab:params} and parameter correlations in Fig.~\ref{fig:corner}. The posterior distribution centers on a region of parameter space that closely matches the observed eclipse times, eclipse depths, and radial velocities. For our best-fitting solution, we find $\chi^2_\text{ETV}=60.90$, $\chi^2_\text{RV}=16.65$, and $\chi^2_{b(t)}=0.16$. The total $\chi^2=77.71$ is very close to 77, the degrees of freedom in the model, indicating that our solution is a satisfactory fit to the data.

\begin{figure}
\includegraphics[width=0.5\textwidth]{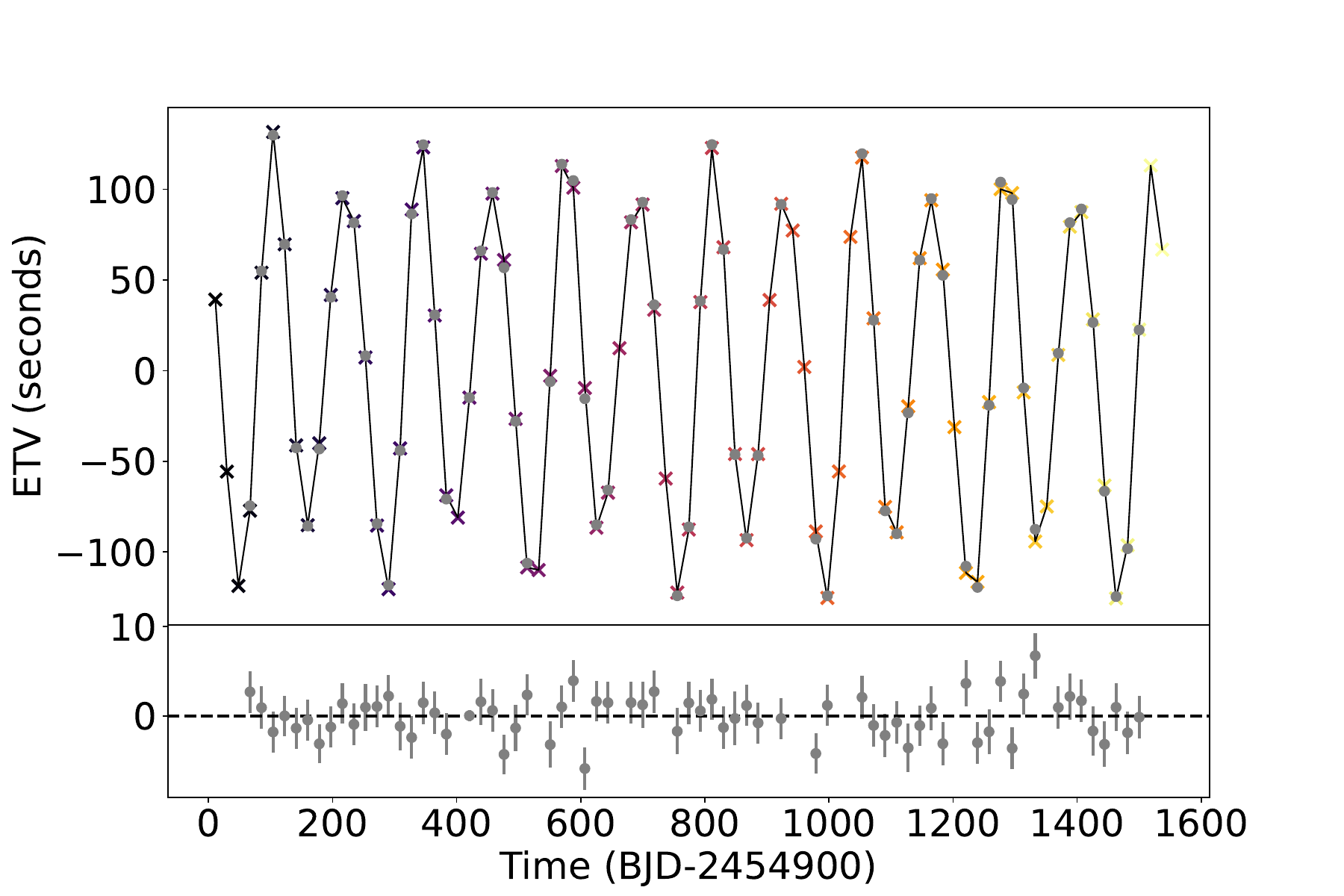}
\caption{The Observed-Calculated primary eclipse times. Eclipse mid-times from the \textit{N}-body model are the colored crosses, times from the light curve fit are the gray circles.}
\label{fig:OMCunfold}
\end{figure}

\begin{figure}
\includegraphics[width=0.5\textwidth]{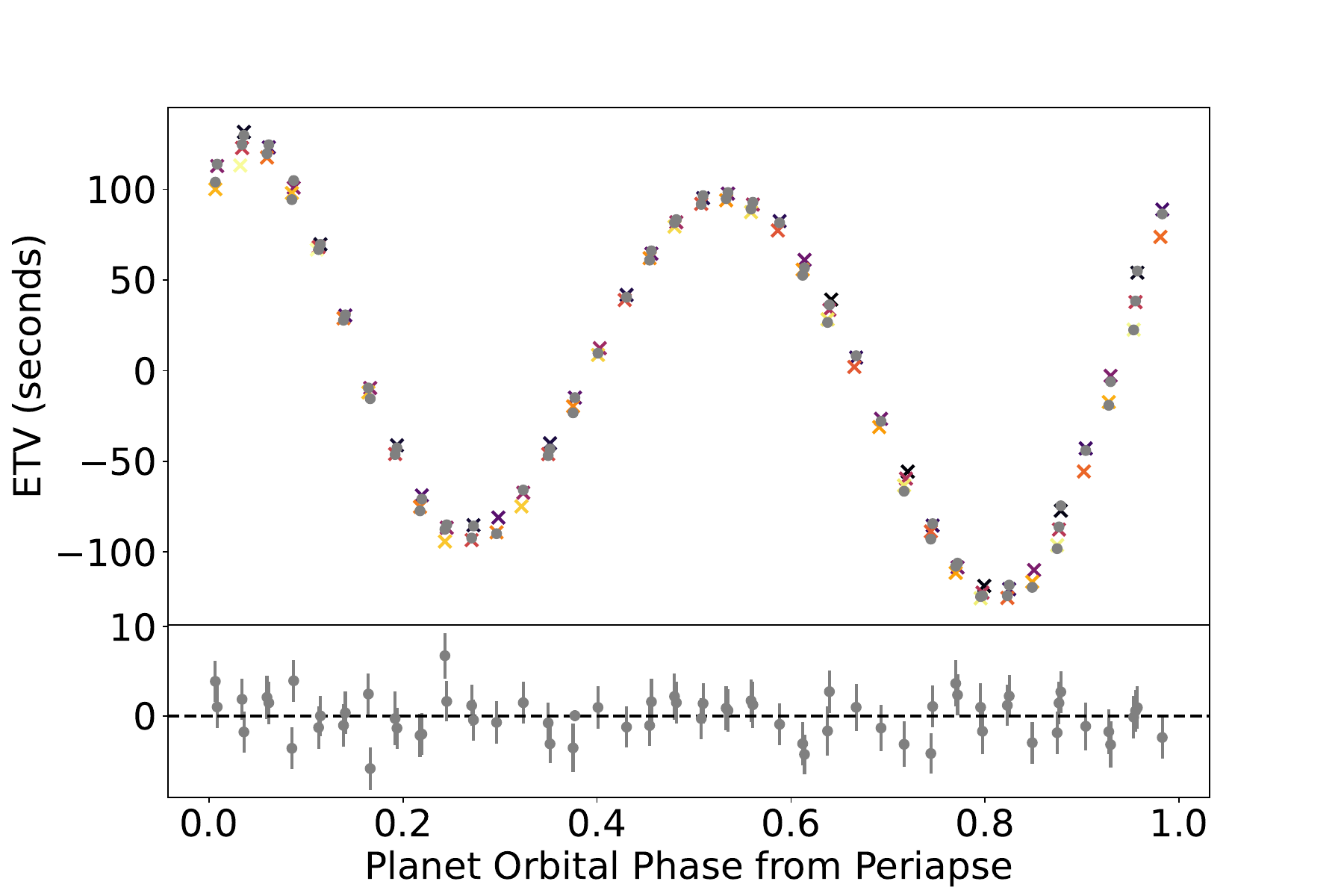}
\caption{Same as Fig.~\ref{fig:OMCunfold} but phase-folded at the peak ETV frequency $235.9$d. Periapse precession of the planet causes the shape of the ETV curve itself to change slightly over the \kepler baseline}.
\label{fig:OMCfold}
\end{figure}

The eclipse time data and our best-fitting solution are shown on an O-C diagram in Fig.~\ref{fig:OMCunfold}. Both the model and data times were subtracted by the same best-fitting linear ephemeris model to emphasize the eclipse timing variations rather than the orbital period. Figure~\ref{fig:OMCfold} contains the same data but phase-folded to the peak Fourier frequency of $235.9$ d. The double-humped structure common to ETVs caused by a third body is apparent, as is a small asymmetry in the widths of the peaks due to the eccentricity of the orbit of the third body. Also visible is a slight shift in the shape of the ETVs over the $\sim 1500$ d observation period. This is because the periapse of the planet, which has a strong effect on the ETV shape \citep{Borkovits2016}, precesses over the \kepler baseline.

\begin{figure}
\includegraphics[width=0.5\textwidth]{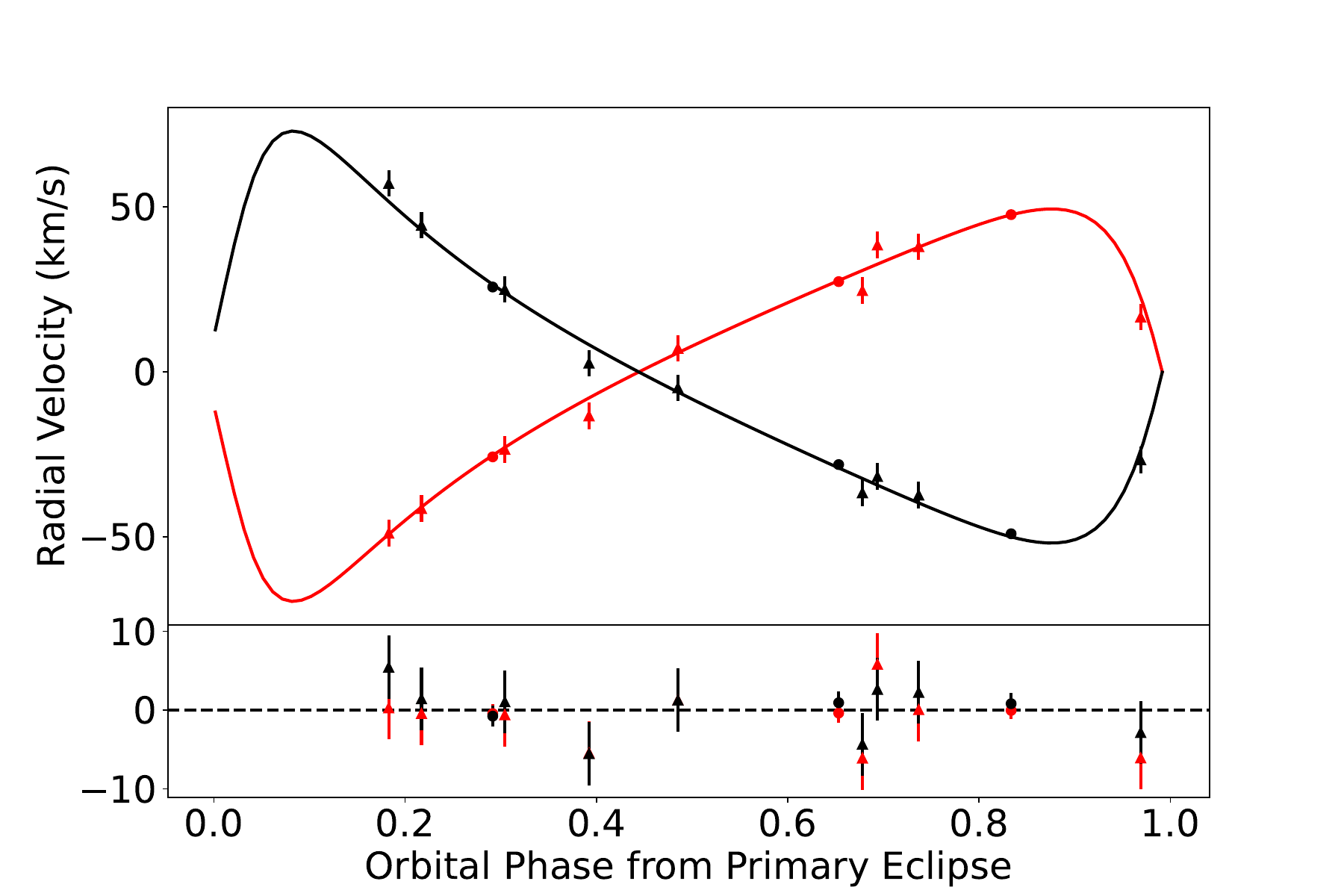}
\caption{Radial velocity observations as a function of binary orbital phase from eclipse. Solid lines are the \textit{N}-body model. Circles are the data from FIES and triangles are the data from CARMENES. The red curve is the primary star and the black curve is the secondary star.}
\label{fig:rv}
\end{figure}

The stellar radial velocities from the dynamical fit are shown in Fig.~\ref{fig:rv}. The non-zero eccentricity and near-unity mass ratio is apparent. The mass determination is hampered by limited phase sampling near the peak radial velocity; nevertheless we determine the stellar masses to better than 10 per cent. In contrast to the results of \cite{Getley2017}, we find the stars to have remarkably similar masses, with $M_A/M_B = 1.05 \pm 0.02$.

\begin{table*}
\caption{The results of the MCMC fit to ETVs, RVs and impact parameter measurements. The orbital elements are instantaneous (osculating) values at the reference epoch $T_\text{ref}=2454900 \text{ BJD}$. \label{tab:params}}
\begin{tabular}{cccc}
Parameter & Maximum Likelihood Value & Median & $1\sigma$ uncertainty \\
\hline
\multicolumn{4}{c}{\textit{Binary Orbit}} \\
\hline
Orbital period, $P_1$ (d) & 18.610875 & 18.610869 & 0.000031 \\
Time of eclipse, $T_{01}$ (BJD-2454900) & 66.861873 & 66.861694 & 0.000302 \\
Eccentricity, $e_1$ & 0.5032 & 0.4973 & 0.0172 \\
Inclination, $i_1$ (deg) & 84.670 & 84.399 & 0.766 \\
Argument of periapse, $\omega_1$ (deg) & 108.34 & 109.09 & 1.79 \\
Ascending node $\Omega_1$ (deg) & 0.0 (fixed) & 0.0 & 0.0 \\
Primary star mass, $M_A$ $(M_\odot)$ & 1.1366 & 1.1450 & 0.0728 \\
Secondary star mass, $M_B$ $(M_\odot)$ & 1.0820 & 1.0980 & 0.0644 \\
\hline
\multicolumn{4}{c}{\textit{Planet Orbit}} \\
\hline
Orbital period, $P_2$ (d) & 239.4799 & 239.5044 & 0.0686 \\
Time of periapse, $T_P$ (BJD-2454900) & 95.811 & 95.730 & 0.815 \\
Eccentricity, $e_2$ & 0.05511 & 0.05541 & 0.00147 \\
Inclination, $i_2$ & 80.17 & 86.31 & 4.21 \\
Argument of periapse, $\omega_2$ (deg) & -46.58 & -46.17 & 1.87 \\
Longitude of ascending node, $\Omega_2$ (deg) & 0.894 & -0.437 & 0.931 \\
Planet mass, $M_p$ $(M_{\rm Jup})$ & 4.885 & 4.992 & 0.230 \\
Mutual inclination, $i_m$ (deg) & 3.35 & 3.71 & 8.21$^a$ \\
\hline
Barycentric radial velocity (NOT), $\gamma_1$ (km s$^{-1}$) & 4.508827 & 4.644869 & 0.550843 \\
Barycentric radial velocity (CARMENES), $\gamma_2$ (km s$^{-1}$) & 77.427103 & 77.442553 & 1.000817 \\
\hline
\multicolumn{4}{l}{$^a$ 95\% upper limit}\\
\multicolumn{4}{l}{}\\
\end{tabular}
\end{table*}

\begin{figure}
    \centering
    \includegraphics[width=0.45\textwidth]{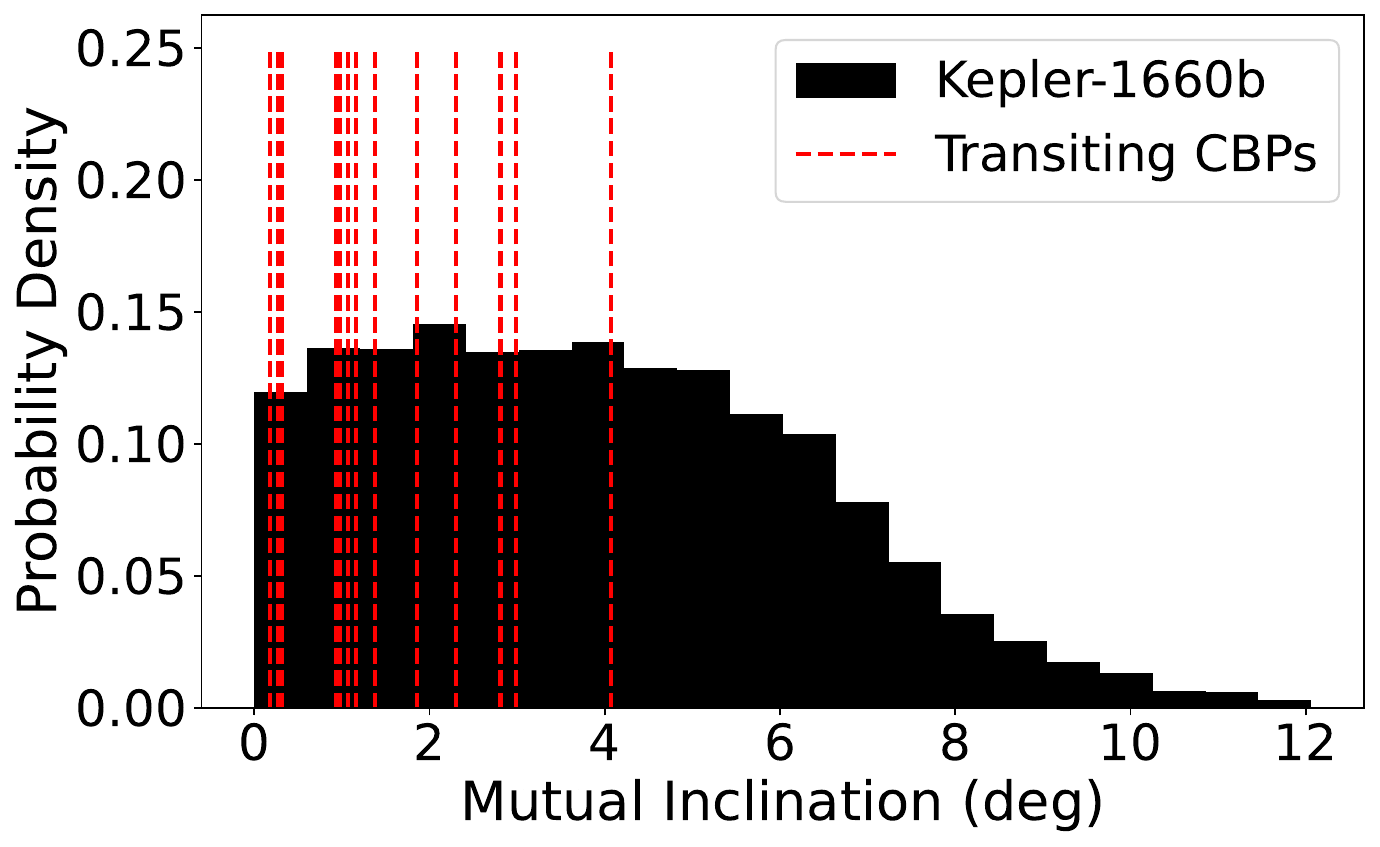}
    \caption{Posterior distribution of the planet-to-binary mutual inclination from our MCMC chains. Dashed red vertical lines mark the best-fit mutual inclinations of the 14 transiting circumbinary planets from \textit{TESS} and \kepler.}
    \label{fig:i_m}
\end{figure}

The dynamical architecture of circumbinary systems depends strongly on the mutual inclination $i_m$ between the binary and circumbinary orbital planes \citep{Farago2010}. This observer-independent value can be computed from angles referenced to the sky-plane by the spherical law of cosines \citep[e.g.][]{Borkovits2015}
\begin{equation}
    \cos{i_m} = \cos{i_1}\cos{i_2} + \sin{i_1}\sin{i_2}\cos{\Delta\Omega}.
\end{equation}
With such a definition, values of $i_m=0^\circ$, $90^\circ$, $180^\circ$ represent prograde coplanar, polar, and retrograde coplanar orbits, respectively.
Figure~\ref{fig:i_m} shows the posterior distribution of the mutual inclination as compared with the transiting circumbinary planet sample \citep{Li2016}. The distribution is consistent with coplanar and we find an 95 per cent certainty upper limit on the mutual inclination of $8.21^\circ$. This coplanarity is consistent with, though less precisely constrained than, the transiting circumbinary planet sample.

\subsection{Constraining alternative solutions}
As noted above, the three descriptions of this system (\citet{Borkovits2016}, \citet{Getley2017}, and this work) differ substantially in their parameters, particularly in the orientation of the planet orbit. To better understand the possible orbital configurations, we mapped slices of the parameter space and determined which regions are consistent with the \kepler eclipse data. Our RV data and the light curve strongly constrain the orbital elements and mass of the binary, therefore we only modify the seven parameters describing the planet orbit and mass and hold the remaining ten fixed to our best-fit solution in Table \ref{tab:params}.

Although we used the sky-plane inclination $i_2$ and relative longitude of ascending node $\Delta\Omega$ to parametrize the planet in the MCMC fit, here we use the more dynamically relevant mutual inclination $i_m$ and dynamical argument of periapse $g_1$. The dynamical argument of periapse, defined in \citet{Borkovits2016}, is the angle between the periapse of the binary orbit and the intersection of the planet orbit with the binary orbital plane.\footnote{The angles $(i_m,g_1)$ reduce to the standard inclination $i$ and longitude of ascending node $\Omega$ in the case that the reference plane and reference direction are the binary orbital plane and line of apsides.} These angles are of particular interest because fixed points in the Hamiltonian of a test particle orbiting an eccentric binary occur in four places: coplanar orbits with $i_m=0^\circ, 180^\circ$, and polar orbits $(i_m, g_1) = (90^\circ, \pm 90^\circ)$. At those points precession of the planet orbital plane vanishes \citep{Farago2010}. Furthermore, circumbinary protoplanetary discs dissipatively evolve into one of these fixed points, making them likely regions for circumbinary discs and planet formation \citep{Martin2017, Kennedy2019}.

First, we show which regions of the $(i_m, g_1)$ parameter space can produce the observed ETVs. At 5000 points uniformly distributed in the circle defined by $0^\circ\leq i_m\leq 180^\circ$, $0^\circ\leq g_1\leq 360^\circ$, we fit the 5 remaining parameters ($P_2$, $T_P$, $e_2\cos\omega_2$, $e_2\sin\omega_2$, $M_p$) to minimize $\chi^2_\text{ETV}$, the sum of squared normalized residuals of the eclipse times.

Second, we show which regions of the $(i_m, g_1)$ parameter space can produced the observed changing eclipse depth. At the same 5000 points in $(i_m, g_1)$ space, we compute the change in impact parameter for the $\chi^2_\text{ETV}$-minimizing solution found previously, and compare it to the measured depth change by calculating $\chi^2_{b(t)}$ (Eq.~\ref{eq:chisq_b}).

\begin{figure*}
    \centering
    \includegraphics[width=\textwidth]{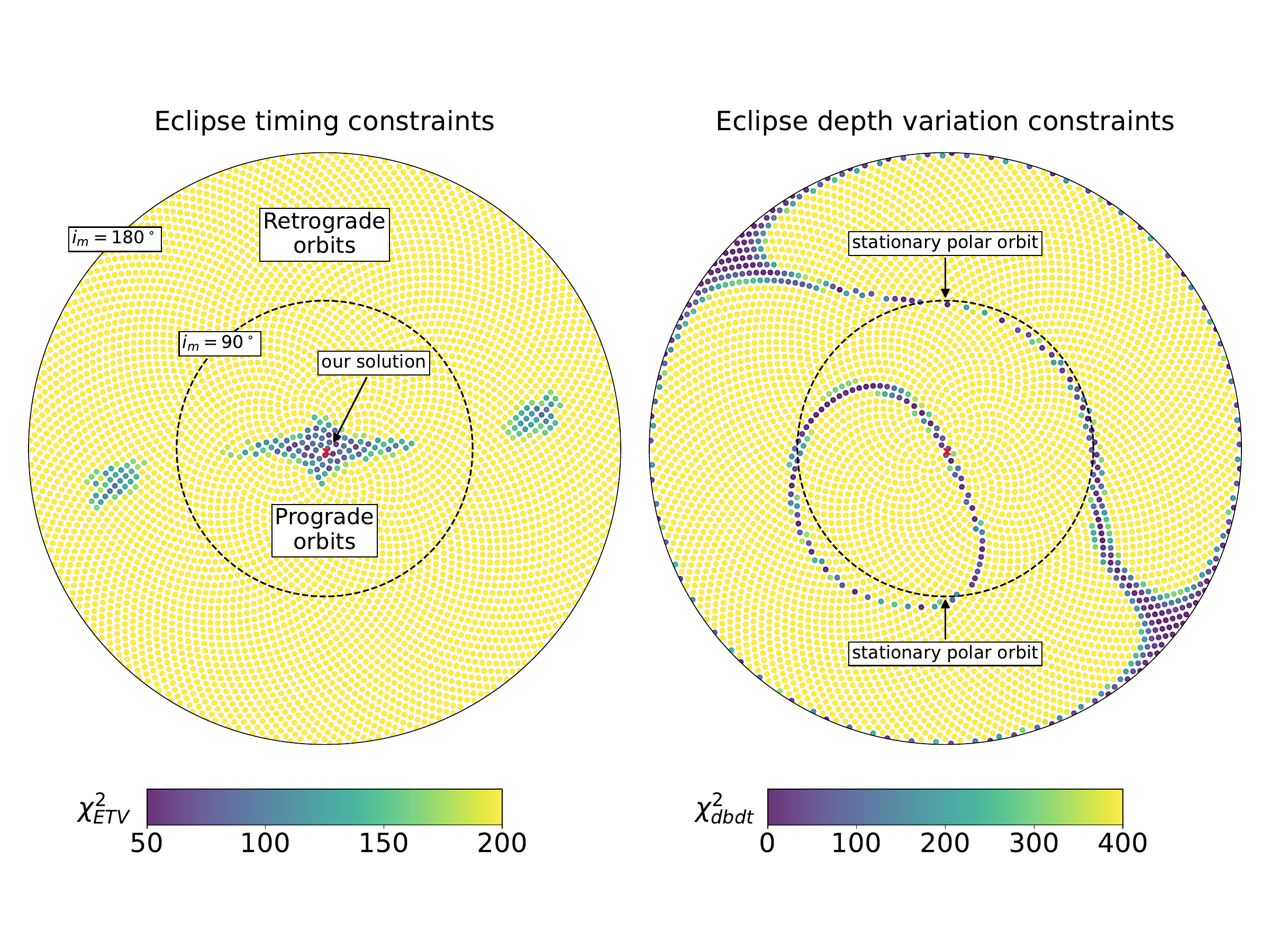}
    \caption{Constraints on planetary orientation from the eclipse timing and eclipse depth changes. On these plots, the radial coordinate is the mutual inclination, $i_m$, and the azimuthal coordinate is the dynamical argument of periapse, $g_1$, defined such that $g_1=0$ on the positive x-axis and increases going counterclockwise. The color of each point on the left panel is the answer to the question, ``for fixed $(i_m, g_1)$ and binary parameters, how well could a planet reproduce the eclipse times?'' Each point on the right panel uses the same best-fitting solution as the respective point on the left but is colored according to how well that solution matches the observed change in impact parameter from the light curve. On both plots, dark dots represent good fits and light dots poor ones, and the red crosses mark the $i_m$ and $g_1$ of Kepler-1660 reported by this work.}
    \label{fig:dbdt}
\end{figure*}

The results are shown in Fig.~\ref{fig:dbdt}. The left plot shows three regions of $(i_m, g_1)$ parameter space that can reproduce the observed eclipse times: a large region centered around $i_m\approx 0^\circ$, and two smaller regions centered around $(i_m, g_1)\approx (130^\circ, 10^\circ)$ and $(130^\circ, 190^\circ)$. However, the right plot demonstrates that these two misaligned configurations are strongly ruled out because they result in depth variations much larger than observed. We can therefore conclude that a prograde and nearly coplanar solution is the only configuration consistent with the RV, eclipse timing, and eclipse depth data.

\section{Discussion}\label{sec:discussion}

\subsection{Discrepancies with previous works}\label{subsec:discrepancies}

The initial analysis of \citet{Borkovits2016} and the follow-up work of \citet{Getley2017} both only modelled the eclipse \textit{timing} variations, and not any eclipse \textit{depth} variations. They also did not have the benefit of radial velocity characterization. Our RV and ETV-only model does allow for some misaligned orbits, including two patches of permissible parameter space with retrograde inclinations similar to the \citet{Getley2017} $i_m = 120^{\circ}$ result. However, the right side of Fig.~\ref{fig:dbdt} demonstrates that by adding in impact parameter constraints, the \citet{Getley2017} solution is ruled out and a coplanar solution is favored. This is reiterated in the mutual inclination posterior plotted in Fig.~\ref{fig:i_m}.

However, our model is in rough agreement with respect to the planet's period (239.48 d vs 237.71 d from \citealt{Getley2017}). Because both works report osculating (instantaneous) orbital elements, and the osculating planet period in our model varies by $\approx 8$ d over the \kepler baseline, the difference is likely attributable to the choice of reference epoch and other orbital parameters. Our mass is lower ($4.89M_{\rm Jup}$ vs $7.698M_{\rm Jup}$ from \citealt{Getley2017}), but still qualitatively a massive `super-Jupiter'. We also note that the \citet{Getley2017} best-fitting binary inclination is $80^{\circ}$, which would not correspond to an eclipsing binary. 

Our work also benefits from our follow-up RVs, which the previous studies did not have at the time but encouraged. These RVs indicate a significantly different set of binary parameters. The clear signature of a double-lined spectroscopic binary reveals a near twin binary, with $M_{\rm A} = 1.14 M_\odot$ and $M_{\rm B} = 1.08 M_\odot$, as compared to the unequal masses from \citet{Getley2017} of $M_{\rm A} = 1.21 M_\odot$ and $M_{\rm B} = 0.51 M_\odot$. We also find a binary that is twice as eccentric, with $e_1=0.50$, compared with \citet{Getley2017}'s 0.246. Discrepancies in the stellar and binary parameters may account for their larger planet mass. Radial velocity characterization is particularly important for binaries such as Kepler-1660AB where there are no secondary eclipses, and hence very little information is known about the eccentricity and mass ratio from photometry alone. However, we note that even though RVs were essential for constraining the planet's parameters, a highly misaligned orbit can be ruled out from the photometry alone, through the lack of eclipse depth variations.

\subsection{Future observations}

\begin{figure}
    \centering
    \includegraphics[width=0.47\textwidth]{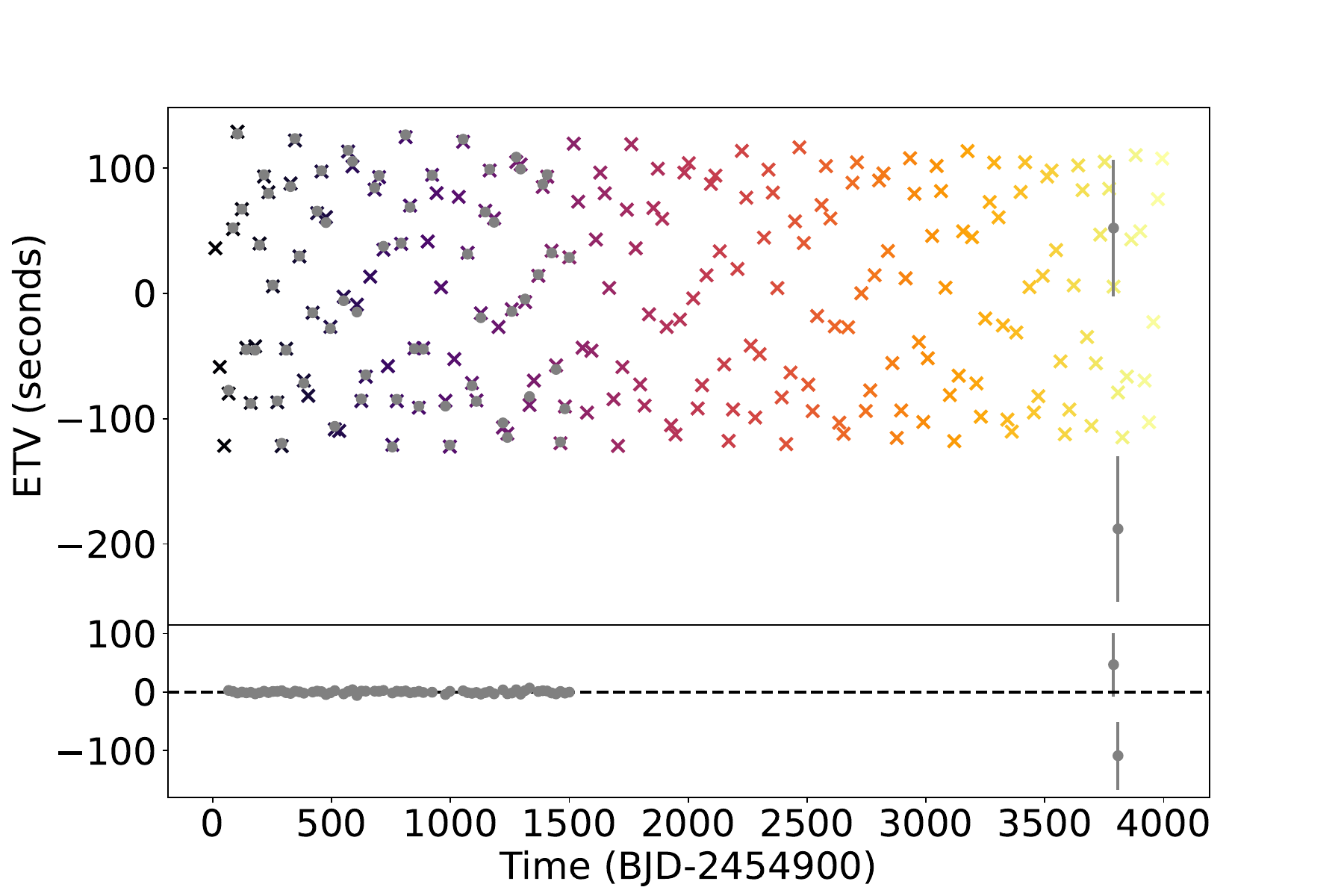}
    \caption{Model ETVs (coloured crosses) and observed ETVs (gray circles). The initial approximately four years of ETVs come from Kepler, with very small error bars. TESS has observed two eclipses, with ETVs matching the model, albeit with very large error bars. The colour of the model ETVs scales with the eclipse time.}
    \label{fig:tess}
\end{figure}

The circumbinary planet Kepler-1660ABb could be independently detected in radial velocities  using a long time series of higher precision radial velocity measurements which are sensitive to the signal of not only the binary, but of the planet itself. This was recently demonstrated with the RV detection of the transiting Kepler-16ABb \citep{Triaud2022} and the discovery of BEBOP-1c/TOI-1338ABc in radial velocities \citep{Standing2023}. 

The expected planet RV semi-amplitude for Kepler-1660ABb is $K=94$ m s$^{-1}$. Our present data are not sufficient to detect the planet mainly because our precision is on the order of km s$^{-1}$. We also only have 12 observations (equal to the number of free parameters with two Keplerian orbits) spanning 206 d (less than the 239.5 d orbit). Finally, Kepler-1660ABb is a double-lined spectroscopic binary, for which spectral contamination makes spectroscopic analysis trickier than the single-lined binaries Kepler-16ABb and  BEBOP-1c/TOI-1338ABc.

An alternative way of detecting the planet is in transit. Even though the planet is currently not in transitability, the nodal precession of its orbit will change its sky orientation over time, potentially bringing it in and out of transitability. This was first proposed by \citet{Schneider1994}, studied in more detail by \cite{Martin2014,Martin2015,Martin2017a} and seen observationally for Kepler-413 \citep{Kostov2014} and Kepler-453 \citep{Welsh2015} discoveries. The period of this circumbinary nodal precession was derived by \cite{Farago2010} as a function of orbital elements and stellar masses. Using their expression we obtain $P_\text{prec}=86$ yr for Kepler-1660.

\cite{Martin2015} derived an analytic criterion for the minimum mutual inclination needed for transitability to occur on the primary (A) or secondary (B) star at some point during the nodal precession cycle:
\begin{equation}
\label{eq:transitability}
    i_{m,\text{min}} > \left|\frac{\pi}{2} - i_1 \right| - \sin^{-1}\left(\frac{a_{\rm A,B}}{a_{\rm p}} \sin \left|\frac{\pi}{2} - i_1 \right| + \frac{R_{\rm A,B}}{a_{\rm p}} \right).
\end{equation}
For Kepler-1660ABb, $i_{m,\text{min}}=3.8^{\circ}$ for guaranteed transitability. Based on our current knowledge, it is not certain that the planet will ever transit. The main reason is that the binary, despite eclipsing, is $\sim 5^{\circ}$ off an exactly edge on orbit. Extending this slightly tilted orbital plane out to the distance of the planet, a planet in this plane (i.e. perfectly coplanar) would not transit. Our 95 per cent certainty upper limit to $i_m$ is $8.2^{\circ}$. Using Eq.~\ref{eq:transitability} we find the probability of the planet ever transiting either to be 70 per cent when marginalized over our posterior distribution. 

Based on a cursory look of existing TESS data, we do not see any transits. TESS has discovered two transiting circumbinary planets so far \cite{Kostov2020,Kostov2021} but its short one-month observing sectors present a challenge for the typically long-period circumbinary planets \citep{Kostov2020a}. The eclipses of the binary are visible in TESS and could provide an opportunity to dramatically enhance the observational baseline of this system. We have determined the mid-times for two such eclipses and found them to be consistent with our model (Figure~\ref{fig:tess}) However, the far reduced photometric performance of TESS compared to \kepler means that these new data do not meaningfully improve our results.

\subsection{Orbital stability}

\begin{figure*}
    \centering
    \includegraphics[width=0.8\textwidth]{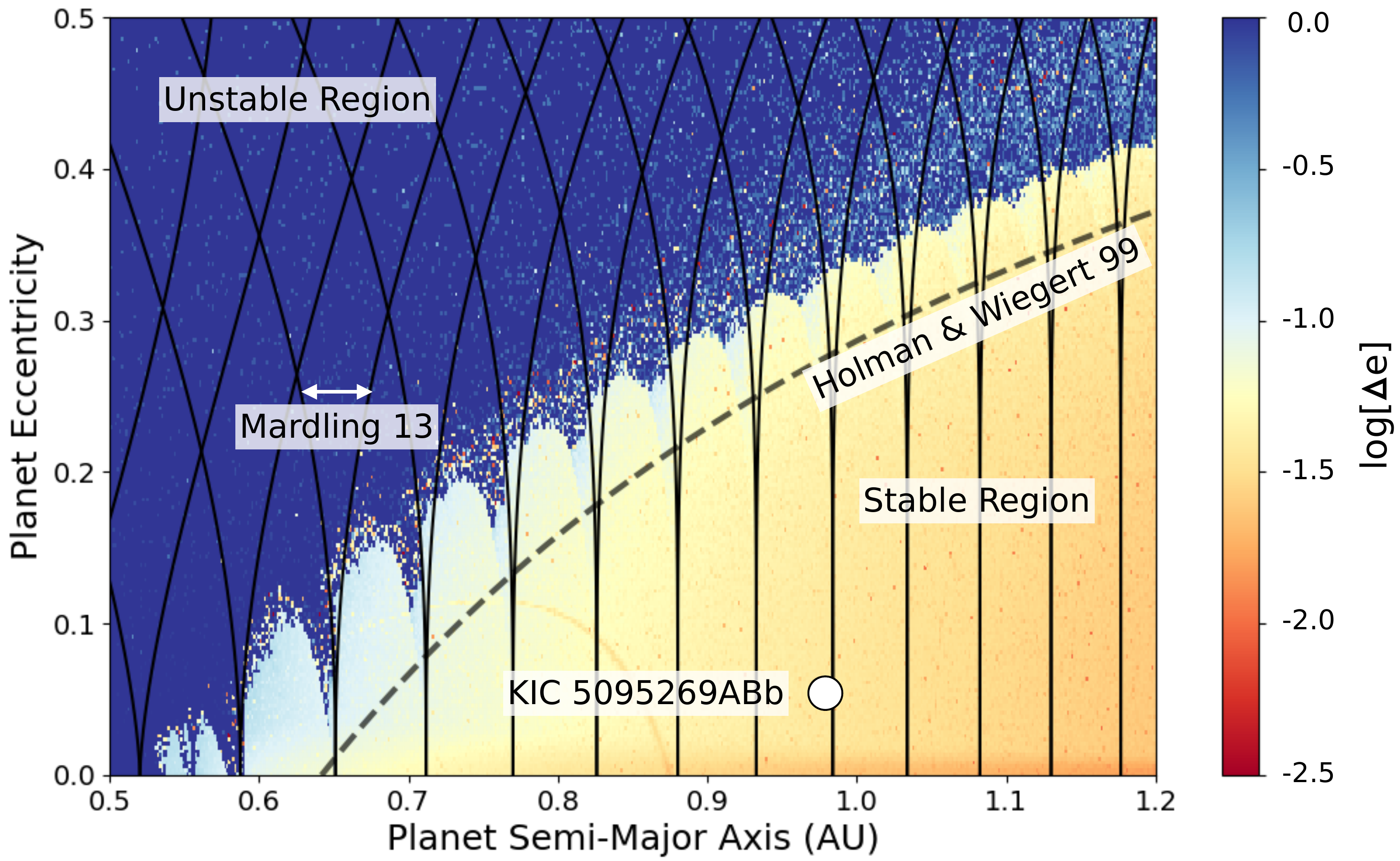}
    \caption{Stability map of planets of different orbital distance and eccentricity around the Kepler-1660AB binary. The color scale indicates $\log(\Delta e)$, the variation in eccentricity over an \textit{N}-body integration time of 5000 years. Blue orbits correspond to unstable planets ejected from the system. The black solid line cones are the mean motion resonant widths from \citet{Mardling2013}, which tend to correspond to heightened eccentricity variation. The black and white dashed line corresponds to the traditional \citet{Holman1999} stability limit, including an empirical scaling added by \citet{Quarles2018} to account for planet eccentricity. The white circle indicates the our derived parameters for the circumbinary planet, safely in a stable parameter space.}
    \label{fig:stability}
\end{figure*}

To confirm that this planet is dynamically stable, we produce a custom `stability map' in Fig.~\ref{fig:stability} using the argument of periapse, period, eccentricity, and masses of the binary from Table \ref{tab:params}. This map shows the planet's eccentricity variation ($\log\Delta e$) for a given starting planet eccentricity and semi-major axis over an evolution time of 5000 yr. The color represents the eccentricity variation, which is indicative of stability, with bluer regions being unstable and redder regions being stable. The overlaid dashed line is the classic stability limit from \citet{Holman1999}, modified by \cite{Quarles2018} to account for the planet's eccentricity. The black wedged lines are the resonant widths from \citet{Mardling2013} and indicate how resonances can create regions of instability at very low eccentricities for lower semi-major axis values. The circumbinary planet in this system, at a semi-major axis just below 1 au with an eccentricity of 0.055, is indicated by the white circle. At this position, the planet sits comfortably in a stable orbit and further evolution will not threaten this stability. 

The planet is farther away from the stability limit than most of the transiting circumbinary planets, which are typically found `piled-up' near where the circumbinary disc inner edge would have been \citep{Martin2018}. Like the transit method, but unlike the light travel time effect, detection by dynamically-induced ETVs will be biased towards planets closer to the binary \citep{Borkovits2015}. Ultimately, a population of circumbinary planets found with different methods will be needed to rigorously assess how piled up planets are near the stability limit. That distribution has significant implications for their formation history of circumbinary planets, as it is believed that they form in the outer regions of the disc before migrating in and parking near the disc edge \citep{Pierens2013,Penzlin2021,Martin2022,Fitzmaurice2022}. 

Nevertheless, as an individual planet, Kepler-1660ABb fits with the \citet{Pierens2008} prediction that massive circumbinary planets should {\it not} be found near the stability limit. Their argument is that the parking of inwards-migrating small planets near the inner disc edge is caused by a significant density bump in the disc. If the planet is very massive (roughly Jupiter or higher) then it will open up a gap in the disc and move inward due to Type-II migration. The gap eliminates the inner pressure bump, shutting off the parking mechanism. The planet then migrates too close to the binary and gets ejected or sent on a much wider orbit. For Kepler-1660ABb to exist on its current orbit likely means the disc dissipated before it could migrate dangerously too close.

\subsection{Formation and existence of circumbinary planets on misaligned orbits}

The initial orbits for Kepler-1660 proposed by \citet{Borkovits2016,Getley2017} were highly misaligned to the binary, by $40^{\circ}$ and $120^{\circ}$ respectively. We emphasize that we rule out these orbits in favour of a coplanar one solely on the basis of the data (in particular the lack of eclipse depth variations). Our conclusion is independent of any questions of the plausibility of misaligned circumbinary orbits. Indeed, there are arguments that they may exist, as we discuss briefly here.

All of the known transiting circumbinary planets have orbits aligned to within $4^{\circ}$ of the host binary orbit (\citealt{Martin2019} and Fig.~\ref{fig:i_m}). The mutual inclination distribution can be roughly described by a Rayleigh distribution with $\sigma=1.5^{\circ}$ \citep{Martin2019b}. However, \citet{Martin2014} showed that transits for a misaligned orbit would be aperiodic and hard to detect, and hence this flat inclination distribution may be an observational bias. More recent work of \citet{Chen2022} comes to a similar conclusion. A flat inclination distribution would imply a circumbinary gas giant frequency similar to that around single stars ($\approx 5-10$ per cent, \citealt{Martin2014,Armstrong2014}). The BEBOP radial velocity survey has not found a large `hidden' population of misaligned planets \citep{Martin2019a,Standing2022}, despite demonstrating that circumbinary planets can be found via radial velocity monitoring of binary stars \citep{Triaud2022,Standing2023}.

It is possible that even if the majority of circumbinary planets are coplanar, exceptional cases of misaligned orbits may exist. There are in fact multiple arguments for the existence of misalignments. From an orbital stability perspective, \citet{Pilat-Lohinger2003,Farago2010,Doolin2011,Martin2016} demonstrated that circumbinary planets may be stable at all mutual inclinations, including retrograde orbits. We have observed circumbinary discs that are misaligned (e.g. KH15D, \citealt{Winn2004,Chiang2004}) and polar (99 Herculis, \citealt{Kennedy2012}, and HD 98800, \citealt{Kennedy2019}). Theory suggests that protoplanetary discs may form with an initially isotropic distribution but reach equilibrium states that are either coplanar or polar \citep{Martin2017,Martin2019}. Terrestrial planet formation is theoretically possible in such polar discs \citep{Childs2021}, although recent work by \citet{Childs2022} suggests that formation in a misaligned disc may lead to coplanar planets if the disc lacks gas. 

Even if a planet formed in a fairly flat orbit, it is possible that planet-planet scattering could place it onto a misaligned orbit \citep{Chatterjee2008}, although there is a risk that scattering may increase not only inclination but eccentricity, potentially ejecting the planet in a binary \citep{Smullen2016,Fitzmaurice2022}. Misaligned orbits may also be produced via secular interactions with an external stellar perturber \citep{Martin2015a}.

Overall, while the orbit of Kepler-1660ABb is not misaligned to its binary, such a system is far from inconceivable and we encourage continued searches for one.

\section{Conclusion}\label{sec:conclusion}

We confirm the existence of a coplanar circumbinary planet Kepler-1660ABb, the first confirmed non-transiting \kepler circumbinary planet. It is a `super-Jupiter', with $M_{\rm p}=4.89M_{\rm Jup}$. It was first identified using ETVs by \cite{Borkovits2016} and later analysed by \citet{Getley2017}. Whereas these two studies proposed a highly misaligned planetary orbit (roughly $40^{\circ}$ and $120^{\circ}$ of misalignment, respectively), we demonstrate that the planet is in fact coplanar (to within $\approx 8^{\circ}$). Our analysis improves upon previous works by accounting for both eclipse {\it timing} variations and eclipse {\it depth} variations, the latter of which rule out a misaligned orbit because there is negligible depth variation. We also include follow-up radial velocities which we took, which allow us to properly measure the stellar masses and binary eccentricity.

Kepler-1660ABb is the most massive circumbinary planet known, but it turns out not to be the first on a significantly misaligned orbit.

\section*{Acknowledgments}
We thank the anonymous referee for comprehensive suggestions that significantly improved this work. This work was completed in part with resources provided by the University of Chicago Research Computing Center, with nodes purchased using the Sloan Research Fellowship. Based on observations made with the Nordic Optical Telescope, owned in collaboration by the University of Turku and Aarhus University, and operated jointly by Aarhus University, the University of Turku and the University of Oslo, representing Denmark, Finland and Norway, the University of Iceland and Stockholm University at the Observatorio del Roque de los Muchachos, La Palma, Spain, of the Instituto de Astrofisica de Canarias. The data from the CARMENES instrument were collected at the previous Centro Astron\'omico Hispano Alem\'an (CAHA; current name: Centro Astron\'omico Hispano en Andaluc\'\i a). Support for DVM was provided by NASA through the NASA Hubble Fellowship grant HF2-51464 awarded by the Space Telescope Science Institute, which is operated by the Association of Universities for Research in Astronomy, Inc., for NASA, under contract NAS5-26555.

\section*{Data Availability}
The eclipse and radial velocity data are included in the article appendix. Posterior samples will be shared on reasonable request to the corresponding author.

\vspace{5mm}

\bibliography{main} \bibliographystyle{mnras}

\appendix

\section{}

\begin{table*}
\caption{Eclipse mid-times from our analysis of the \kepler light curve.}
\label{tab:ecl}
\begin{tabular}{ccc}
Eclipse Index & Eclipse Time (BJD-2454900) & $1\sigma$ uncertainty (d)\\
\hline
 -40  &    66.8651676  &   0.0000271 \\
 -39  &    85.4786256  &   0.0000278 \\
 -38  &   104.0914523  &   0.0000267 \\
 -37  &   122.7027159  &   0.0000264 \\
 -36  &   141.3133773  &   0.0000265 \\
 -35  &   159.9248384  &   0.0000266 \\
 -34  &   178.5372897  &   0.0000253 \\
 -33  &   197.1502183  &   0.0000268 \\
 -32  &   215.7628265  &   0.0000261 \\
 -31  &   234.3746132  &   0.0000271 \\
 -30  &   252.9857246  &   0.0000306 \\
 -29  &   271.5966131  &   0.0000270 \\
 -28  &   290.2081810  &   0.0000272 \\
 -27  &   308.8210001  &   0.0000311 \\
 -26  &   327.4344698  &   0.0000276 \\
 -25  &   346.0468713  &   0.0000276 \\
 -24  &   364.6577455  &   0.0000276 \\
 -23  &   383.2685300  &   0.0000272 \\
 -21  &   420.4930974  &   0.0000075 \\
 -20  &   439.1059939  &   0.0000299 \\
 -19  &   457.7183260  &   0.0000279 \\
 -18  &   476.3298060  &   0.0000259 \\
 -17  &   494.9407861  &   0.0000298 \\
 -16  &   513.5518408  &   0.0000265 \\
 -14  &   550.7769189  &   0.0000301 \\
 -13  &   569.3902668  &   0.0000275 \\
 -12  &   588.0021224  &   0.0000273 \\
 -11  &   606.6126901  &   0.0000273 \\
 -10  &   625.2238455  &   0.0000261 \\
 -9  &   643.8360273  &   0.0000271 \\
 -7  &   681.0616729  &   0.0000270 \\
 -6  &   699.6737434  &   0.0000302 \\
 -5  &   718.2850490  &   0.0000278 \\
 -3  &   755.5071126  &   0.0000301 \\
 -2  &   774.1195116  &   0.0000272 \\
 -1  &   792.7329129  &   0.0000273 \\
  0  &   811.3458716  &   0.0000269 \\
  1  &   829.9571616  &   0.0000274 \\
  2  &   848.5678133  &   0.0000348 \\
  3  &   867.1792408  &   0.0000267 \\
  4  &   885.7917271  &   0.0000265 \\
  6  &   923.0172498  &   0.0000268 \\
  9  &   978.8509935  &   0.0000262 \\
 10  &   997.4625919  &   0.0000267 \\
 13  &  1053.3012928  &   0.0000278 \\
 14  &  1071.9121903  &   0.0000279 \\
 15  &  1090.5229339  &   0.0000279 \\
 16  &  1109.1347480  &   0.0000280 \\
 17  &  1127.7474789  &   0.0000312 \\
 18  &  1146.3604138  &   0.0000263 \\
 19  &  1164.9727660  &   0.0000284 \\
 20  &  1183.5842367  &   0.0000284 \\
 22  &  1220.8063003  &   0.0000300 \\
 23  &  1239.4181253  &   0.0000273 \\
 24  &  1258.0312472  &   0.0000291 \\
 25  &  1276.6446312  &   0.0000269 \\
 26  &  1295.2564795  &   0.0000270 \\
 27  &  1313.8672377  &   0.0000267 \\
 28  &  1332.4782956  &   0.0000297 \\
\end{tabular}
\end{table*}

\begin{table*}
\contcaption{}
\begin{tabular}{ccc}
Eclipse Index & Eclipse Time (BJD-2454900) & $1\sigma$ uncertainty (d)\\
\hline
 30  &  1369.7033394  &   0.0000277 \\
 31  &  1388.3161322  &   0.0000301 \\
 32  &  1406.9281795  &   0.0000280 \\
 33  &  1425.5394157  &   0.0000316 \\
 34  &  1444.1502984  &   0.0000297 \\
 35  &  1462.7615862  &   0.0000311 \\
 36  &  1481.3738519  &   0.0000279 \\
 37  &  1499.9872078  &   0.0000278 \\
\end{tabular}
\end{table*}

\begin{table*}
\caption{Our radial velocity data. All velocities are given in km s$^{-1}$, and errors are $1\sigma$ uncertainties.}
\label{tab:RVs}
\begin{tabular}{c c c c c c}
Time (BJD-2454900) & Primary RV & Primary RV Error & Secondary RV & Secondary RV uncertainty & Telescope \\
\hline
3087.429909  &  -21.1  &  1.2  &  30.3  &  1.3 & NOT/FIES \\
3131.385773  &   32.0  &  1.2  & -23.5  &  1.4 & NOT/FIES \\
3153.353683  &   52.3  &  1.1  & -44.4  &  1.4 & NOT/FIES  \\
3299.647071 &   116.4  &  4.0 &  46.3   &  4.0 & CARMENES \\
3312.647215 &   64.6  &   4.0 &  80.6   &  4.0 & CARMENES \\
3329.620515 &   54.5  &   4.0 &  103.0  &  4.0  & CARMENES \\
3336.578951 &   102.6  &  4.0 &  41.3   &  4.0 & CARMENES \\
3346.610853 &   36.6  &   4.0 &  122.4  &  4.0 & CARMENES \\
3351.600984 &   85.1  &   4.0 &  73.1   &  4.0 & CARMENES \\
3360.605952 &   94.6  &   4.0 &  51.3   &  4.0 & CARMENES \\
3364.585143 &   29.1  &   4.0 &  135.1  &  4.0 & CARMENES \\
3393.510143 &   115.9  &  4.0 &  40.6   &  4.0 & CARMENES \\
\end{tabular}
\end{table*}

\begin{figure*}
\includegraphics[width=\textwidth]{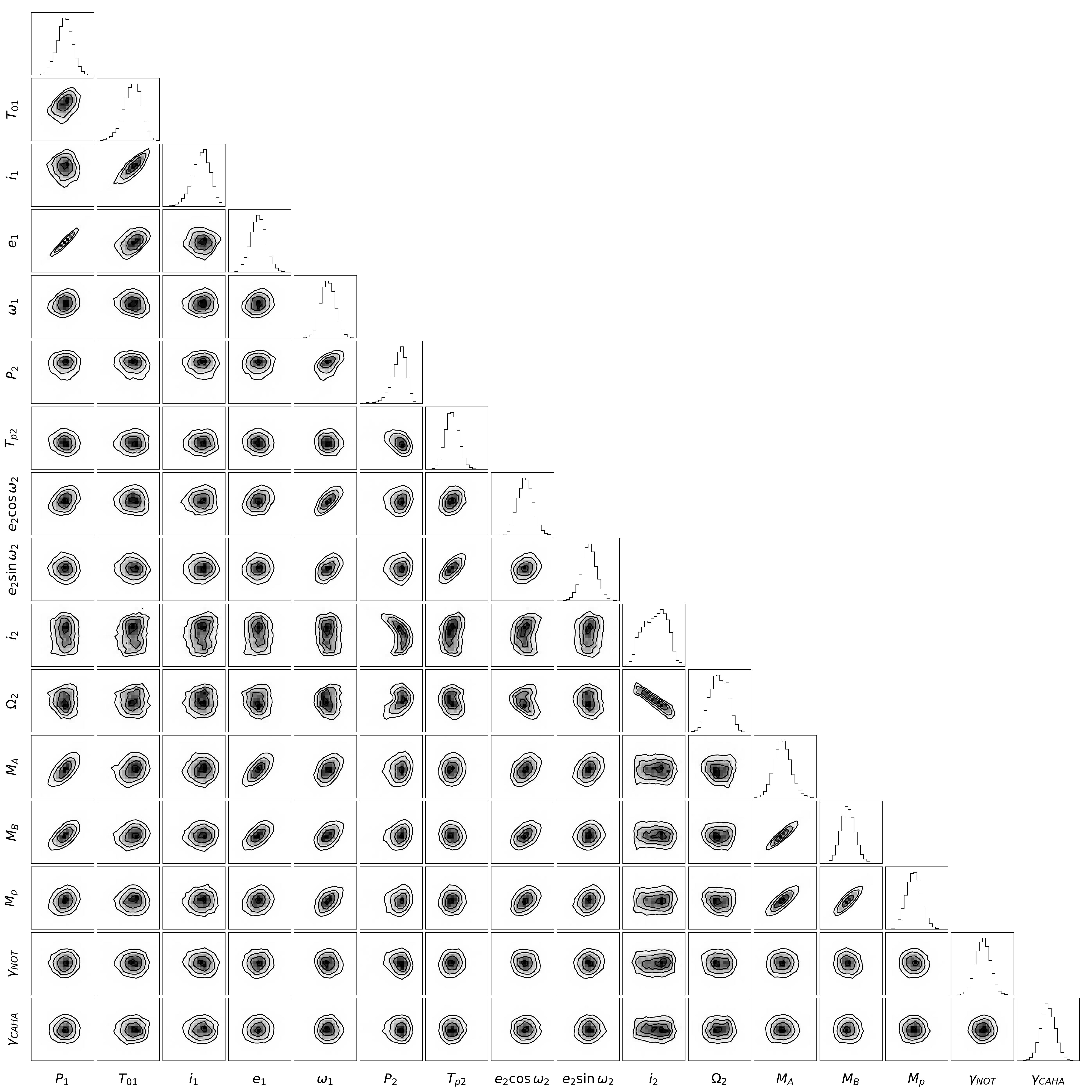}
\caption{Scatter plot of correlations between the parameters of our model. Contour levels mark 0.5, 1, 1.5, and 2$\sigma$ bounds. See the main body for the precise parameter definitions and values.}
\label{fig:corner}
\end{figure*}

\bsp	
\label{lastpage}
\end{document}